# Laser-Architected Surface Wetting

Lara Tomholt[1,2], Forrest Meggers[2,3], Reza Moini[1]

[1] Department of Civil and Environmental Engineering, Princeton University, Princeton, NJ, USA

[2] Andlinger Center for Energy and the Environment, Princeton University, Princeton, NJ, USA

[3] School of Architecture, Princeton University, Princeton, NJ, USA

**Abstract**

Technologies that require surface wetting or evaporative cooling require the ability to efficiently spread fluids across large areas, as increased wetted surface area increases evaporative flux. However, the intrinsic surfacial and bulk properties of most engineered materials substantially limit the rate and magnitude of surface wetting and lack control of flow direction, preventing them from rapidly wetting large surfaces. Here, we introduce our approach for rapid and controlled wetting of surfaces by laser-engraving channel networks that provide pathways for rapid, long-distance (cm-dm scale) capillary fluid propagation across the area, while the intrinsic material properties enable slow, short-distance (mm-cm scale) surface wetting. We investigated this approach on hardened cement paste and showed that laser engraving is a fabrication-friendly, scalable, and reproducible solution for creating channels with properties conducive to capillary fluid propagation. We demonstrate that the rate and direction of surface wetting can be controlled by tuning the channel network density, channel network anisotropy, and supplied fluid flow rate. The integration of laser-engraved channel networks demonstrated significantly greater wetting performance (up to 10-fold greater wetted area and up to 180-fold greater wetting performance when wetted area is adjusted for fluid use efficiency) and greater evaporative cooling (up to 1.8 °C cooler surfaces) compared to control (hardened cement paste without laser-engraved channel networks).

## Introduction

Surface wetting and evaporative cooling technologies require the ability to efficiently spread fluids across large areas, as increased wetted surface area increases evaporative flux [1]. However, the intrinsic surfacial and bulk properties of most engineered materials (e.g., surface energy, surface roughness, and pore microstructure) limit the rate and magnitude of surface wetting, and therefore limit the flow rates they can accommodate, and lack the potential for controlling the direction of flow, preventing them from rapidly wetting large surfaces [2,3]. Nature provides interesting solutions to overcome this limitation. Biological systems such as plant leaves [4] and elephant skin [5], leverage larger capillary channels for rapid, longer-distance fluid propagation, arranged in an interconnected network that optimizes distribution across the area, in combination with shorter fluid pathways through permeable tissues for slower, local wetting. Drawing on these biological strategies, we present a unique solution for rapid and controlled wetting of large surfaces, first introduced in July 2023 [6], that employs laser-engraved channel networks to provide pathways for rapid, long-distance fluid propagation across the area (driven by a combination of a single inlet flow rate, capillary action, and gravity), while the intrinsic material properties enable slow, short-distance surface wetting (Fig. 1, Fig. 2A). Here, we introduce laser engraving as a highly precise, reproducible, and effective method to create open-groove capillary channels into hardened cement paste. Furthermore, we investigated the effect of the channel network design on the direction of flow, surface wetting rate, and fluid use efficiency, particularly on vertical surfaces where flow is influenced by gravity. We subsequently examined the effect of the increased surface wetting capabilities of these laser-engraved surfaces on evaporative cooling performance.

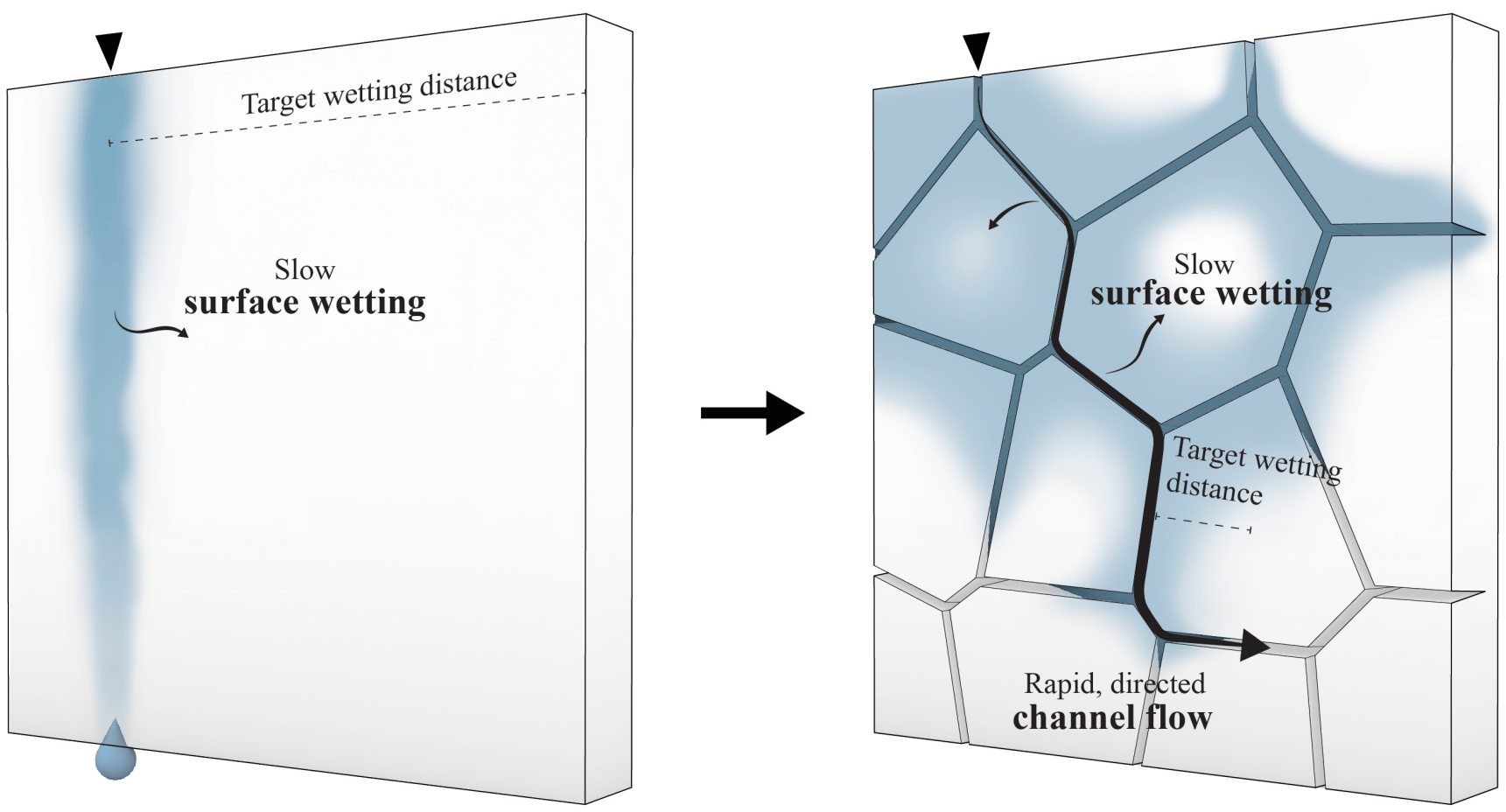

**Fig. 1.** Limited and slow surface wetting in monolithic material vs. rapid and extended wetting in laser-engraved counterparts: The intrinsic surfacial and bulk properties of a material limit imbibition rates, and consequently, allow only for limited and slow wetting. This prevents extensive wetting of large surfaces, limits the fluid flow rate the material can accommodate, and consequently results in significant amounts of fluid loss from the system (left diagram). In contrast, by creating channel networks into the surface (right diagram), fluid pathways are created that allow for rapid and directed fluid propagation across the area (driven and controlled by a combination of a single inlet flow rate, capillary action, and gravity), while also reducing the target propagation distance for the slow wetting of the material. These surface modifications significantly increase the wetting performance and water use efficiency.

## Results

We demonstrated that the laser-engraved channels are highly reproducible and have ideal properties for capillary-driven flow, including a v-shape cross-section with a small free perimeter (less than 500 µm), a high depth-to-width aspect ratio (~ 8:1), and hydrophilic channel walls (contact angle of 16.5°) (Fig. 2A). By controlling the laser traveling speed and laser power, channels of different dimensions can be created (Fig. 2BC). We analytically calculated how channel dimensions control the speed of capillary flow, the maximum flow distance, and the time the channel first overflows under varying flow directions. Numerical simulations demonstrated that the 'bridges' that partially enclose the channels, created by the melting and resolidifying of the hardened cement paste during the laser engraving, further enhance capillary fluid propagation.

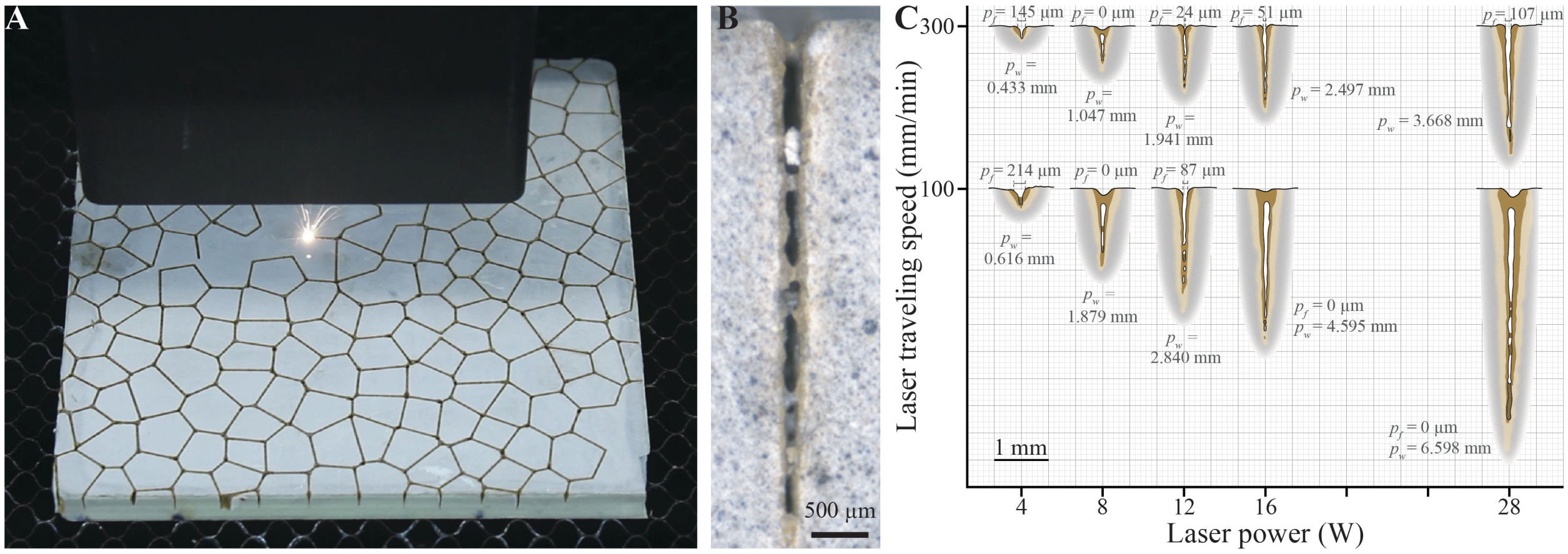

**Fig. 2.** Capillary channels enabled by laser-engraving: **A.** Laser-engraving the surface of a hardened cement paste sample with a laser cutter. **B.** Cross-section of a laser-engraved channel into hardened cement paste, using a laser power of 28 W and a laser traveling speed of 100 mm/min. **C.** The effect of the laser power and laser traveling speed on the shape and dimensions of the resulting channel. Here, $p_w$ is the wetted perimeter and $p_f$ is the free perimeter of the representative channel section displayed.

We examined fluid flow and adjacent surface wetting in single channels (200 mm long) and on 100 x 100 mm vertical surfaces (10 mm thick). For these experiments, we designed and built a custom experimental setup with controlled environmental conditions (23 ± 0.9 °C, 45 ± 1.9 %RH, fluid and samples at environment temperature) that supplies each sample with water at a fixed flow rate and simultaneously images surface wetting, records the surface temperature, measures the weight of the sample being wetted, and measures the weight of the water dripping from the sample.

We demonstrated how the rate and distance of fluid propagation through a laser-engraved channel and the time of channel overflow can be controlled by tuning the inlet flow rate and channel inclination. We also show that wetting of the surface adjacent to the channel occurs within one second of the advancing fluid front in the channel. We demonstrated how wetting of the vertical surfaces with water can be carefully controlled by tuning the channel network density (i.e., Chebyshev radius of the cells) (Fig. 3), anisotropy of the network design (Fig. 3), and inlet flow rate. More efficient channel networks were able to accommodate the high supply flow rate (5 $mm^3$/s), recapturing and redirecting water from overflowing channels, resulting in a roughly 10x greater wetted surface area within 20 min than hardened cement paste without engraved channel networks (Fig. 4ABC). With a sample size of n=3, we were able to demonstrate the reproducibility of the direction and extent of wetting.

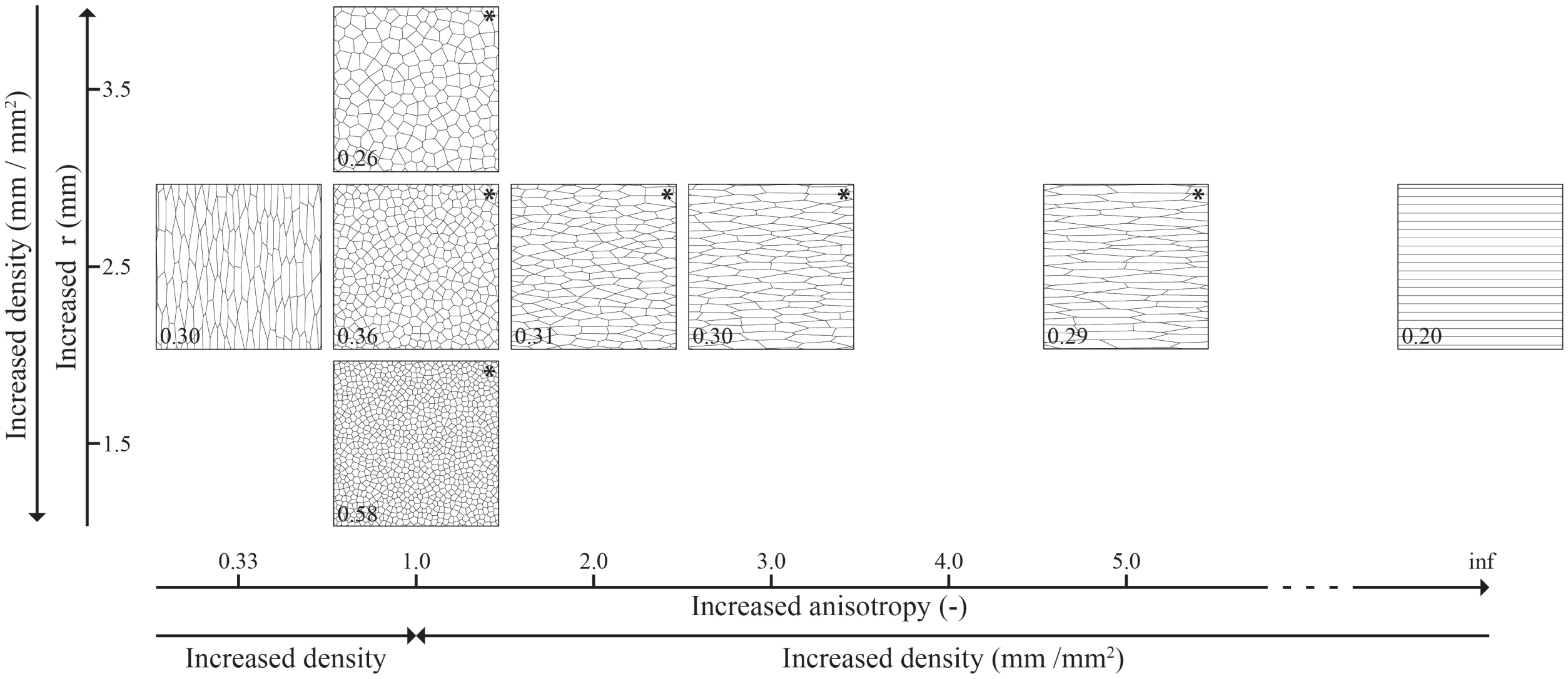

**Fig. 3.** The main channel networks designed for experimental testing of surface wetting and evaporation rate: the voronoi networks have either a fixed anisotropy and different Chebyshev radius of the cells (vertical axis) (e.g., anisotropy = 1.0, and Chebyshev radius of 1.5 mm, 2.5 mm, and 3.5 mm), or a fixed Chebyshev radius and different anisotropy (horizontal axis) (e.g., Chebyshev radius = 2.5 mm, and anisotropy of 0.33, 1.0, 2.0, 3.0, 5.0, and infinity). Channel network density is listed on each pattern, * indicates networks tested with an n=3.

While greater channel network densities (i.e., smaller Chebyshev radius of the cells) reduced the wetting rate, in channels with an anisotropy = 1 they reached larger wetted surface areas at $t$ = 60 min (Fig.4D). Greater anisotropy of the channel network increased both the wetted area of the channel networks tested (from 36% to 96% of the surface at $t$ = 20 min for the lowest and highest anisotropy tested, respectively) (Fig. 4BCE) and the water use efficiency (from 0.50 to 0.96 at $t$ = 20 min for the lowest and highest anisotropy tested, respectively). With the wetting performance adjusted for water use efficiency, hardened cement paste with engraved channel networks achieved up to 180x greater performance than hardened cement paste without engraved channel networks. Increased supply flow rates increased the wetting rate (Fig. 4F), but reduced the water use efficiency, resulting in an optimal flow rate of 5 $mm^3$/s (for the channel network and flow rates tested) for the greatest wetting performance adjusted for water use efficiency.

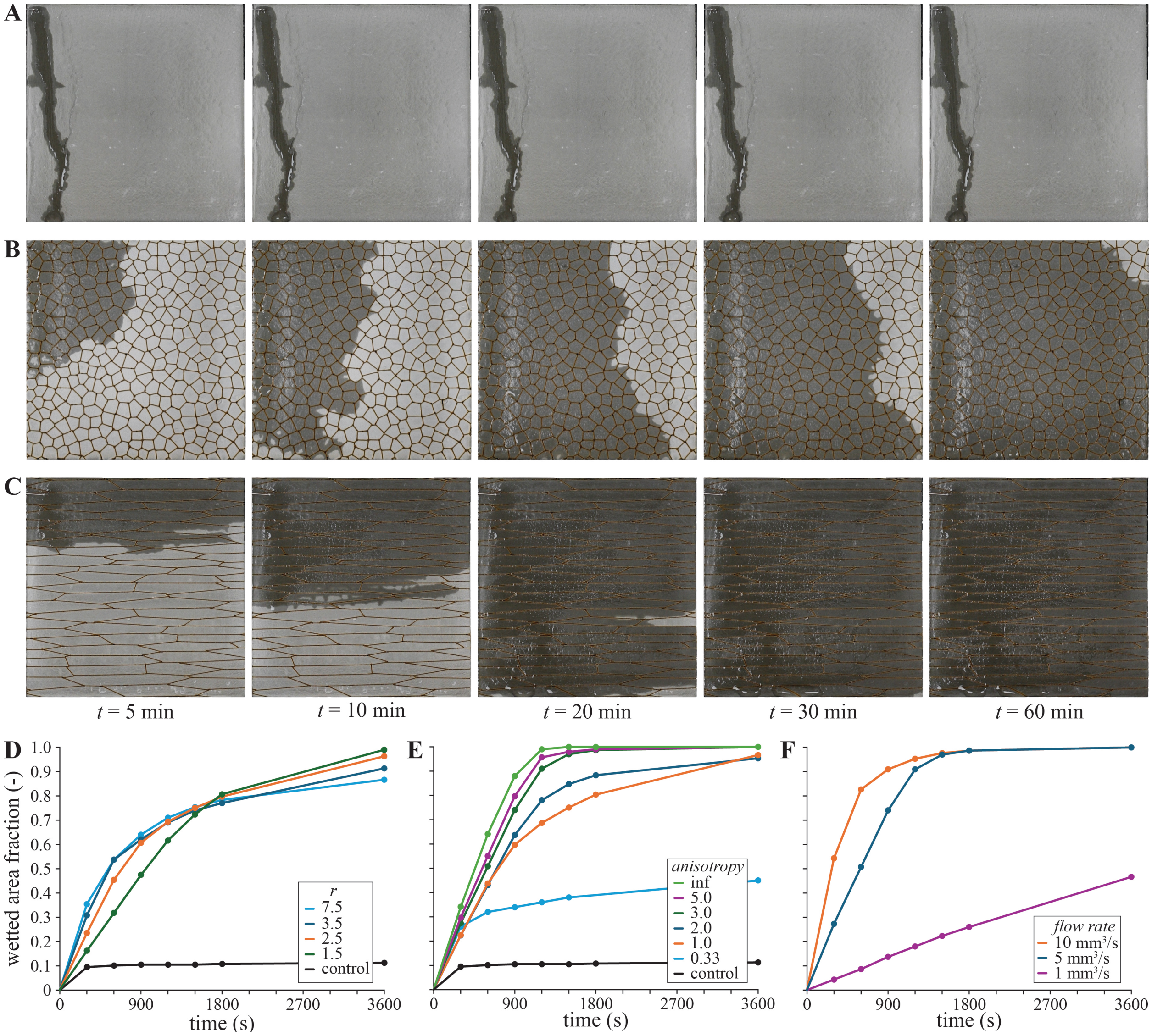

**Fig. 4.** Selected results from the surface wetting experiments, clearly demonstrating that the laser-engraved channel networks enable extensive wetting of the surface by providing rapid long-distance fluid transport and controlling the direction of flow (fluid is supplied from the top left corner). **A.** Hardened cement paste without laser-engraved channels (control). **B.** Hardened cement paste with laser-engraved channels in a network with anisotropy = 1.0 and Chebyshev radius = 2.5 mm. **C.** Hardened cement paste with laser-engraved channels in a network with anisotropy = 5.0 and Chebyshev radius = 2.5 mm. The experiments shown in (A, B, C) were performed with an inlet flow rate of 5 $mm^3/s$. **D.** The effect of the network cells' Chebyshev radius on the wetted area over time for channel networks with an anisotropy = 1.0. **E.** The effect of the network anisotropy on the wetted area over time for channel networks with a Chebyshev radius = 2.5. **E.** The effect of supply flow rate on the wetted area over time for channel networks with a Chebyshev radius = 2.5 and network anisotropy = 3.0.

We additionally illustrated one potential benefit of this rapid surface wetting capability: under our experiments' environmental conditions and with a supply flow rate of 5 $mm^3/s$, evaporation from the most extensively wetted surface cooled the surface roughly 1.8 °C more than hardened cement paste without integrated channel networks at $t$ = 60 min (Fig. 5). Greater temperature reductions can be expected with larger samples and in hotter and drier environments. Infrared imaging showed spatiotemporal cooling patterns that closely match surface area wetting patterns. Our results demonstrated a clear inverse correlation between the integral of the wetted surface area with respect to time and the surface temperature, as well as between the total mass of evaporated water and the surface temperature.

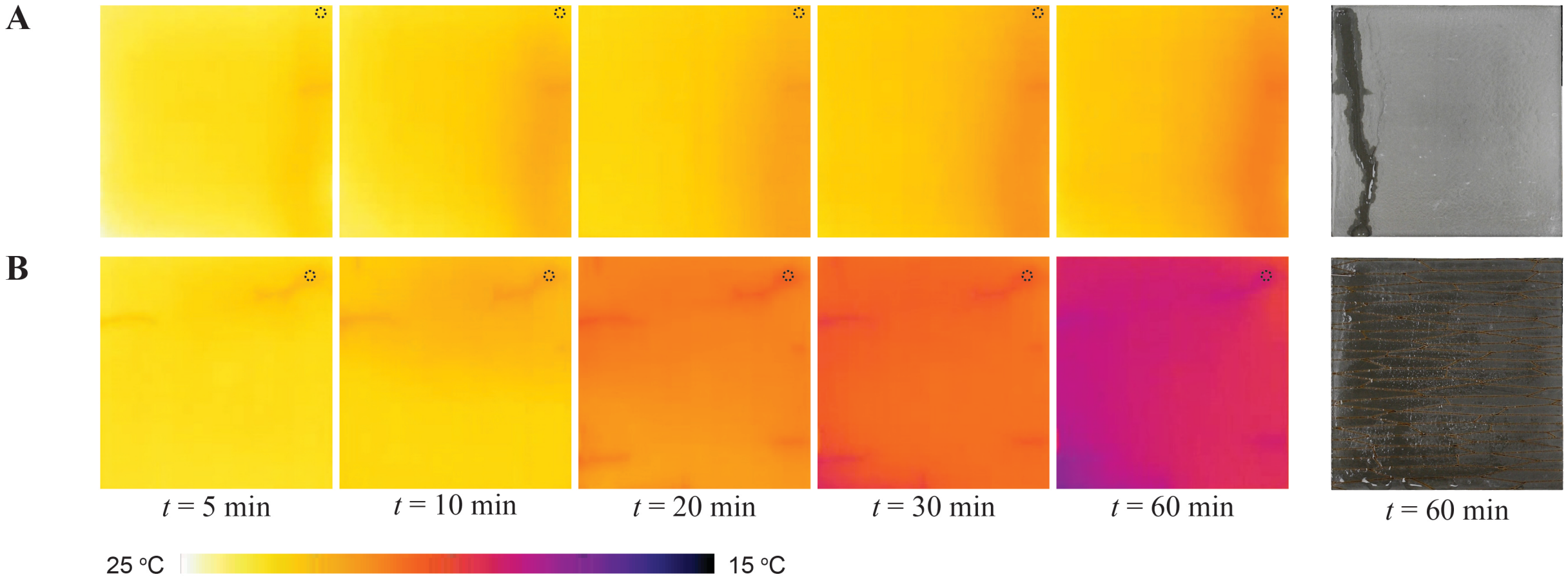

**Fig. 5.** Surface temperature measurements of the back of the sample in two of the experiments performed during our research. Fluid is supplied with an inlet flow rate of 5 $mm^3/s$ on the front of the sample on the top left corner (top right corner in the infrared images, marked with dotted circle). The results clearly demonstrate a significant reduction in surface temperature by more extensive surface wetting, enabled by lasered-engraved channel networks. **A.** Hardened cement paste without laser-engraved channels (control). **B.** Hardened cement paste with laser-engraved channels in a network with anisotropy = 5.0 and Chebyshev radius = 2.5 mm.

## Conclusion

Rapid surface wetting across large areas of engineered materials is rather challenging due to the intrinsic surfacial and bulk properties limiting the magnitude and rate of surface wetting and lack the potential for controlling the direction of flow. The research demonstrated a novel and effective approach to rapidly propagate fluid across the surface of hardened cement paste by integrating a network of laser-engraved channels that provide pathways for rapid, long-distance (cm-dm scale) fluid propagation and, consequently, effective surface wetting.

The research highlighted that laser-engraving hardened cement paste is a fabrication-friendly, scalable, and reproducible solution for creating channels with properties conducive to capillary fluid propagation (including a sharp v-shape cross-section with a small free perimeter, a high depth-to-width aspect ratio, and hydrophilic channel walls). We demonstrated that the rate and direction of surface wetting can be controlled by tuning the channel network density, channel network anisotropy, and fluid supply flow rate. The integration of laser-engraved channel networks demonstrated significantly greater wetting performance, fluid use efficiency, and evaporative cooling compared to hardened cement paste without laser-engraved channel networks.

The proposed method of laser engraving capillary channel networks is scalable to larger panel assemblies, making it viable for large-scale applications such as evaporative cooling solutions for building facades and infrastructure systems. Its general concept of leveraging a network of larger capillary channels to create preferential pathways, propagate fluid over larger distances, and accommodate larger flow rates can also be applied to other types of porous materials or non-porous materials with enhanced surface wettability (e.g., coatings, micro-structured surfaces). Its general concept of integrating larger capillary channels in materials with limited wetting capabilities may change how we design wetting technologies that aim for rapid and extensive surface wetting but traditionally only leverage intrinsic material properties for fluid propagation.

**Acknowledgment**

The team would like to thank the Andlinger Center for Energy and Environment for the Fellowship that supported this research from 2022 to 2024, enabling us to explore novel solutions for surface wetting and actively pursue this research. This work was further supported by Princeton University's Intellectual Property Accelerator Fund (2024) [7]. We would also like to thank Princeton University's Office of Technology Licensing and New Ventures for their guidance and support with the submission of our Invention Disclosure (November 27, 2023) and the filing of a provisional patent (January 3, 2025) [8].